\documentclass[12pt,aps,prb,preprint]{revtex4-1}
\usepackage{graphicx}
\usepackage{url}

\begin{document}


\title{Lorentz-invariant three-vectors and alternative formulation of  relativistic dynamics}


\author{Krzysztof R\c ebilas}
\email[]{krebilas@ar.krakow.pl}
\affiliation{Zak\l{}ad Fizyki. Uniwersytet Rolniczy im. Hugona Ko\l{}\l{}\c ataja w Krakowie. 
Al. Mickiewicza~21, \mbox{31-120} Krak\'ow. Poland.
}


\date{\today}

\begin{abstract}
Besides the well known scalar invariants, there exist also vectorial invariants in the realm of special relativity.  It is shown that the three-vector $\left(\frac{d\vec{p}}{dt}\right)_{\parallel v}+\gamma_v\left(\frac{d\vec{p}}{dt}\right)_{\perp v}$
 is invariant 
under the Lorentz transformation. The indices $_{\parallel v}$ and $_{\perp v}$ denote the respective components established with respect to the
 direction of the  velocity of body $\vec{v}$, and $\vec{p}$ is the relativistic momentum. We prove that this vector is equal to a force of $\vec{F}_R$ satisfying the classical Newtonian law
 $\vec{F}_R=m\vec{a}_R$ in the instantaneous
 inertial rest frame of an accelerated body.  Therefore   the equation 
$\vec{F}_R=\left(\frac{d\vec{p}}{dt}\right)_{\parallel v}+\gamma_v\left(\frac{d\vec{p}}{dt}\right)_{\perp v}$, based on the Lorentz-invariant vectors, may be used as a truly invariant (not merely a covariant) relativistic equation of motion in any inertial system of reference. An alternative approach to classical electrodynamics based on the invariant three-vectors is proposed.
\vspace{1cm}

\noindent
Copyright (2010) American Association of Physics Teachers. This article may be downloaded for personal use only. Any other use requires prior permission of the
 author and the American Association of Physics Teachers.
The following article appeared in Am. J. Phys. 78, 294 (2010) and may be found at \url{http://ajp.aapt.org/resource/1/ajpias/v78/i3/p294_s1}.

\end{abstract}
\maketitle
\section{Introduction}
The origins of the principle of relativity are usually attributed to Galileo, \cite{GAL} who used his famous illustration of a
 moving ship to show that none of the experiments performed on the ship ''below decks'' allows us to decide whether the ship is at
 rest or in steady motion. The same idea was later expressed by Newton \cite{NEW} as the statement that the movements of bodies ''are
 the same among themselves''  regardless of whether the space in which the bodies are placed is at rest or  moves uniformly forward in a straight line. 
Both Galileo and Newton referred to laws of classical mechanics. Their statements were generalized to all laws of physics
 by Poincar\'{e} \cite{POI} and Einstein, \cite{EIN} and the contemporary formulation of 
  the principle of relativity states that the laws of physics have the same form in all inertial reference frames. In Newtonian dynamics, for forces independent of velocity and depending only on some {relative} distances between bodies, this principle is satisfied by Newton's second law of motion:
\begin{equation}
\vec{F}=m\vec{a}
\label{FA}
\end{equation}
because acceleration $\vec{a}$ and a function $\vec{F}$ representing force are \emph{invariant vectors} under Galilean transformation of 
coordinates $\vec{r}\,'=\vec{r}-\vec{V}t$. Invariance of the law follows, then, from  invariance of the vectors this law encompasses. 
Acceleration and force have the status of absolute quantities in Newtonian mechanics, so the law (\ref{FA}) also remains absolutely
 valid regardless of  the inertial frame in which it is used.

In special relativity dynamics, the principle of relativity states that 
 the laws of physics must be invariant under the Lorentz transformation. As we know, the three-dimensional relativistic equation of motion is:
\begin{equation}
\vec{F}=\frac{d\vec{p}}{dt}\;\;,
\label{EQ1}
\end{equation}
where $\vec{p}$ is the relativistic momentum defined as $\vec{p}=m\gamma_v \vec{v}$ with $\gamma_v=(1-v^2/c^2)^{-1/2}$. However, contrary to acceleration in classical  mechanics, $d\vec{p}/dt$ is \emph{not} invariant under the Lorentz transformation. The rate of change of momentum $ d\vec{p}/{dt}$ 
in a frame $S$ is related to the rate of change of momentum $ d\vec{p'}/{dt'}$ in another frame $S'$ that has a velocity 
$\vec{V}$ in the frame $S$ (see Fig. 1) 
\begin{figure}[h!t]
\center{\includegraphics*[width=8cm, ]{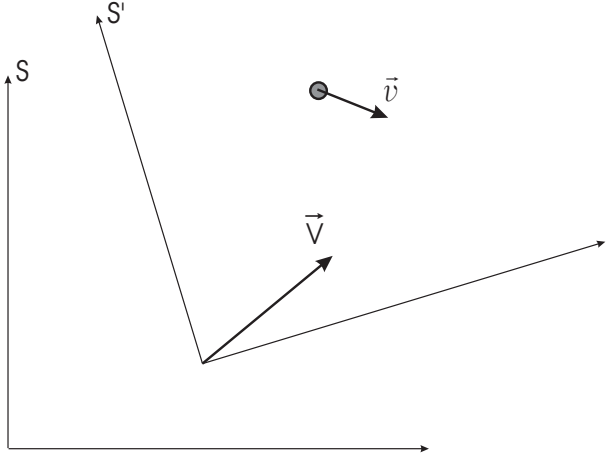}}
\caption{The frame $S'$ moves with a velocity $\vec{V}$ with respect to the frame $S$. A velocity of a body measured in the frame $S$ is $\vec{v}$. Note that the orientation of the axes of  the frames $S$ and $S'$ can be chosen completely arbitrarily because in all our three-dimensional equations the components of the vectors are established with respect to the velocity $\vec{V}$. \label{FIG1}}
\end{figure}
by the equality: \cite{J,KR}
\begin{equation}
\frac{d\vec{p}}{dt}= \left(\frac{d\vec{p'}}{dt'}\right) _{\parallel V} + \gamma_{_ V} \left(\frac{d\vec{p'}}{dt'}\right) _{\perp V} +  \gamma_{_ V}\vec{v} \times \left(\vec{\frac{V}{c^2}} \times  \frac{d\vec{p'}}{dt'}  \right),
\label{pp'}
\end{equation}
where the indices $_{\parallel V}$ and  $_{\perp V}$ refer to the directions parallel and perpendicular to the velocity $\vec{V}$. (Note that the vector $\vec{v}$ on the right side of this equation is the velocity of a body measured in the frame $S$). The postulate of relativity is then fulfilled in the realm of relativity in some other sense. The equation of motion is written by means of four-vectors:
\begin{equation}
\frac{d{p}^\mu}{d\tau}=K^\mu,
\label{41}
\end{equation}
where $(p^\mu)=(m\gamma_vc,\vec{p})$, $\tau$ is a scalar invariant of the Lorentz transformation called the proper time, and $(K^\mu)=(K_0,\vec{K})$ is a four-vector, of which the spatial part $\vec{K}$ is connected with the
 force $\vec{F}$, as $\vec{K}=\gamma_{_ v}\vec{F}$. Because $(p^\mu)$ and $(K^\mu)$ are four-vectors, the left and right sides of the above equation transform in the same manner: $ p^\mu =\Lambda^\mu_\nu p^{'\nu} $, and $ K^\mu =\Lambda^\mu_\nu K^{'\nu} $, where $\Lambda^\mu_\nu$ are components of a tensor representing a Lorentz transformation (a generalized one, if a rotation of axes is included) from the frame $S'$ to the frame $S$. Inserting these relations into (\ref{41}), and remembering that $\Lambda^\mu_\nu$ are time-independent, Eq.~(\ref{41}) appears to be equivalent to:  
\begin{equation}
\frac{d{p}^{'\nu}}{d\tau}=K^{'\nu}.
\label{42}
\end{equation}
Eq.~(\ref{42}), being the relativistic equation of motion in the frame $S'$, has precisely the \emph{same form} as the relativistic equation of motion (\ref{41}) written in the frame $S$. We say that equation (\ref{41}) is Lorentz-{covariant}. The principle of relativity is thus fulfilled in the sense that \emph{invariance} of the relativistic equation of motion is understood as  its Lorentz-\emph{covariance}.

Let us emphasize  the essential difference between the formulation of the principle of relativity in the realm of relativistic mechanics and its classical meaning. First of all, the four-dimensional equation of motion (\ref{41}) is not \emph{truly} invariant in the sense that it is not based on invariant vectors.  The four-vectors $(p^\mu)$ and $(p^{'\mu})$ are {not} invariant, and likewise neither are the four-vectors $(K^\mu)$ and $(K^{'\mu})$. We have no ''objective'' quantity, analogical to acceleration in Newtonian mechanics, that would represent a numerically invariant (universal) result of the action of force. The invariance of the relativistic law of motion is then merely \emph{structural}, i.e. it is \emph{form-invariant}, as shown in Eqs.  (\ref{41}) and (\ref{42}), but not numerically invariant as it is in the case of the classical law of motion (\ref{FA}).

 Another question concerning the principle of relativity formulated by means of covariant four-vectors is  that Eq.~(\ref{41}) is \emph{redundant}. 
Contrary to four-vector $(t,\vec{x})$ in which all components are relevant for a description of a physical situation, in Eq.~(\ref{41}) physically meaningful is only 
its spatial part  equivalent to Eq.~(\ref{EQ1}). In fact, the zero-components of $(p^\mu)$ and $(K^{\mu})$  are not independent of their spatial parts 
because  $dp^0/dt=(d\vec{p}/dt)\cdot \vec{v}/c$, which means also that $K^0=\gamma_{_ v}\vec{F}\cdot \vec{v}/c=\vec{K}\cdot \vec{v}/c$. In this context, the construction of the four-vectors $(p^\mu)$ and $(K^\mu)$ as representing a physical reality may seem artificial. As the complete
 physical information about the behavior of an accelerating object is included in the spatial part of  $(dp^\mu/ d\tau)$, it is rather strange that what is real in our physical world is to be represented rather by the superfluous  
four-vectors than simply by the fully relevant Euclidean three-vectors $d\vec{p}/dt$ and $\vec{F}$.

However, the advantage of the tensor approach is that it serves to express in an elegant way the requirement of covariance of the law of motion. If we had restricted ourselves only to the three-dimensional formulation of the law in the form  of Eq.~(\ref{EQ1}), the requirement of covariance would have had to be  expressed in a much less concise way,  as  the necessity that the function $\vec{F}$ representing force should transform  covariantly  with the vector $d\vec{p}/dt$, i.e. according to Eq.~(\ref{pp'}):
\begin{equation}
\vec{F}= \left(\vec{F'}\right) _{\parallel V} + \gamma_{_ V} \left(\vec{F'}\right) _{\perp V} +  \gamma_{_ V}\vec{v} \times \left(\vec{\frac{V}{c^2}} \times  \vec{F'}  \right),
\label{FF'}
\end{equation}
The relations (\ref{pp'}) and (\ref{FF'})  entail that the equation $\vec{F'}=d\vec{p'}/dt'$ is satisfied  in the frame $S'$, i.e. we get the three-dimensional equation of motion in the same form as in the frame $S$.

One can verify directly  that the spatial components of Eq.~(\ref{41}) actually transform in accordance with Eqs.~(\ref{pp'}) and (\ref{FF'})  so that both the tensor and the three-vector approach are completely equivalent. Nevertheless, the covariance requirement referring to the fully relevant three-dimensional physical law (\ref{EQ1}) being expressed by means of (\ref{pp'}) and (\ref{FF'})  seems to be much less transparent than its four-vector version. On the other hand, the clarity of the four-vector approach is achieved at the expense of the artificial construction of the  redundant four-vectors $(p^\mu)$ and $(K^\mu)$.

The aim of this paper is to show that besides the known scalar invariants (like the space-time interval or other tensor contractions), there exist also \emph{Lorentz-invariant three-vectors}. Using them, an alternative approach to relativistic dynamics is possible, and a ''strong'' meaning of the principle of relativity - similar to that from Newtonian mechanics - can be preserved.
The cornerstone of our theory is the notion 
that the three-dimensional vector
 $\left(\frac{d\vec{p}}{dt}\right)_{\parallel v}+\gamma_v\left(\frac{d\vec{p}}{dt}\right)_{\perp v}$
 is {invariant} under the Lorentz transformation and that this vector amounts to the force $\vec{F}_R$ established in the instantaneous inertial 
rest frame $R$ of a moving body. Similarly to the Galilean transformation, which leaves acceleration unchanged and allows
  the law of motion $\vec{F}=m\vec{a}$ to be invariant (for forces independent of the velocity of body), the 
vector $\left(\frac{d\vec{p}}{dt}\right)_{\parallel v}+\gamma_v\left(\frac{d\vec{p}}{dt}\right)_{\perp v}$
 may serve as an invariant (absolute) measure 
of the action of force (analogous to $m\vec{a}$) and permits us to 
formulate the truly invariant relativistic equation of
 motion $\vec{F}_R=\left(\frac{d\vec{p}}{dt}\right)_{\parallel v}+\gamma_v\left(\frac{d\vec{p}}{dt}\right)_{\perp v}$ that 
is not only structurally, but also numerically the same in any inertial system of reference.   Contrary to the invariance
 in the realm of Galilean transformation, no restrictions need to be imposed  on force.

 Finally, an alternative approach to classical electrodynamics is presented.

\section{Three-dimensional Lorentz-invariant vector}

Let us refer to Eq.~(\ref{pp'}) and choose the primed frame to be the instantaneous \emph{inertial} \emph{rest} frame  $R$ which can be connected  at any moment with an accelerating body. Then the velocity of the body in the frame $R$ is zero and 
$(d\vec{p}/dt)_R=m\,d\vec{v}_R/d\tau$, where $m$ is the rest mass of the object and $d\vec{v}_R$ is its velocity change measured in the frame $R$. In some other system of reference $S$ the velocity of the body is $\vec{v}$, so that the velocity of the frame $R$ measured in $S$ is $\vec{V}=\vec{v}$ (Fig. 2). 
\begin{figure}[h!t]
\center{\includegraphics*[width=5.5cm, ]{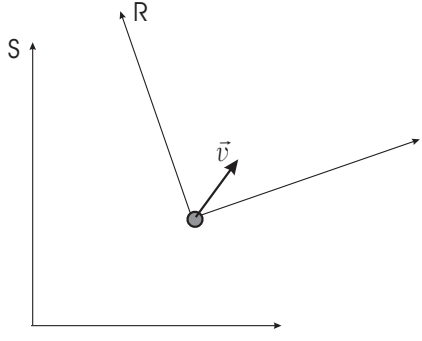}}
\caption{The situation seen from the point of view of the frame S.  \label{FIG2}}
\end{figure}
The relation (\ref{pp'}) then 
 can be rewritten as:
\begin{equation}
{m\frac{d\vec{v}_R}{d\tau}}= \left(\frac{d\vec{p}}{dt}\right) _{\parallel v} + \gamma_{v} \left(\frac{d\vec{p}}{dt}\right) _{\perp v} .
\label{prp'}
\end{equation}
Note that the respective projections refer to the direction determined by the velocity $\vec{v}$. 

On the right hand side of Eq.~(\ref{prp'}) we have only quantities characterizing the moving object as seen in the frame $S$.  
As the frame $S$ is chosen completely arbitrarily, it follows that in \emph{any} inertial  frame the vector $ m\,d\vec{v}_R/d\tau $ can be expressed in the same form  as the vector $\left(\frac{d\vec{p}}{dt}\right)_{\parallel v}+\gamma_v\left(\frac{d\vec{p}}{dt}\right)_{\perp v}$. It means that it is a Lorentz-invariant, and in contrast to the transformation (\ref{pp'}), we have for arbitrary frames $S$ and $S\,'$ the equality:
\begin{equation}
\left(\frac{d\vec{p}}{dt}\right) _{\parallel v} + \gamma_{v} \left(\frac{d\vec{p}}{dt}\right) _{\perp v} =\left(\frac{d\vec{p'}}{dt'}\right) _{\parallel v'} + \gamma_{v'} \left(\frac{d\vec{p'}}{dt'}\right) _{\perp v'}.
\label{pinv}
\end{equation}

To further verify the correctness of our conclusion (\ref{pinv}), let us calculate the contraction of the four-vector $ ({dp^\mu}/{d\tau})$. As we know, such a contraction is a scalar Lorentz-invariant, and to get its value we can calculate it in the rest frame $R$. In result we get:
\begin{equation}
 \frac{dp^\mu}{d\tau}\frac{dp_\mu}{d\tau}=-\left(m\frac{d\vec{v}_R}{d\tau}\right)^2
\label{i1}
\end{equation}
On the other hand we have:
\begin{equation}
\left(\frac{dp^\mu}{d\tau}\right)=\left(\gamma_v \frac{d\vec{p}}{dt}\cdot\vec{\beta}, \gamma_v \frac{d\vec{p}}{dt}\right),
\end{equation}
where $\vec{\beta}=\vec{v}/c$, and we can write:
\begin{equation}
\frac{dp^\mu}{d\tau}\frac{dp_\mu}{d\tau}=\gamma_v^2\left(\beta^2 \left(\frac{d\vec{p}}{dt}\right)^2_{\parallel v}-\left(\frac{d\vec{p}}{dt}\right)^2\right).
\end{equation}
Because $\left(\frac{d\vec{p}}{dt}\right)^2=\left(\frac{d\vec{p}}{dt}\right)^2_{\parallel v}+\left(\frac{d\vec{p}}{dt}\right)^2_{\perp v}$ and $\beta^2-1=-1/\gamma_v^2$, we obtain:
\begin{equation}
\frac{dp^\mu}{d\tau}\frac{dp_\mu}{d\tau}=-\left[\left(\frac{d\vec{p}}{dt}\right)^2_{\parallel v}+\gamma_v^2\left(\frac{d\vec{p}}{dt}\right)^2_{\perp v}\right],
\end{equation}
or
\begin{equation}
\frac{dp^\mu}{d\tau}\frac{dp_\mu}{d\tau}=-\left[\left(\frac{d\vec{p}}{dt}\right)_{\parallel v}+\gamma_v\left(\frac{d\vec{p}}{dt}\right)_{\perp v}\right]^2.
\label{i2}
\end{equation}
The last equation is a direct confirmation that the length of $\left(\frac{d\vec{p}}{dt}\right)_{\parallel v}+\gamma_v\left(\frac{d\vec{p}}{dt}\right)_{\perp v}$ is a Lorentz-invariant and certainly Eqs.~(\ref{i1}) and (\ref{i2}) agree with our prime result (\ref{prp'}).

To better elucidate our idea, let us refer to an analogy with the space-time interval between two events $ds^2=c^2dt^2-d\vec{x}^2$ which is the well known scalar Lorentz-invariant. If the interval is  time-like ($ds^2>0$), one can find such a frame in which $d\vec{x}=0$, i.e. the two events occur at the same place and $ds/c$ represents the \emph{proper} time interval $d\tau$ between the events. Certainly, the value of a proper quantity $d\tau$ established in a rest frame is not affected by the Lorentz transformation, and this is why it can be called Lorentz-invariant. One can thus say that invariance of $ds^2$ derives from invariance of the proper quantity it represents. To put it simply,
 the invariant  $ds^2=c^2dt^2-d\vec{x}^2$ represents one and the same proper quantity $c^2d\tau^2$, merely expressed by means of  coordinates $(t,\vec{x})$ used in a chosen reference frame. 
Similarly, the space-like interval ($ds^2<0$), for which one can find a frame where the two events are simultaneous, is the Lorentz-invariant because it represents an invariant proper length $|d\vec{x}_0|$  of the increment (to be more precise: $ds^2=-|d\vec{x}_0|^2$).  In other words, in any frame, $ds^2=c^2dt^2-d\vec{x}^2$ is simply the proper quantity $-|d\vec{x}_0|^2$, but expressed by coordinates pertaining to the given system of reference. 

In the case of  the vector $\left(\frac{d\vec{p}}{dt}\right)_{\parallel v}+\gamma_v\left(\frac{d\vec{p}}{dt}\right)_{\perp v}$, the situation is quite similar: this vector is Lorentz-invariant because in any frame it is simply {one and the same} \emph{proper} vector ${m\,{d\vec{v}_R}/{d\tau}}$, merely expressed by means of  coordinates connected with a chosen inertial system of reference. Used in the form $\left(\frac{d\vec{p}}{dt}\right)_{\parallel v}+\gamma_v\left(\frac{d\vec{p}}{dt}\right)_{\perp v}$, it can serve as a universal quantity which is \emph{structurally and numerically} the same in any reference frame (analogous to $ds^2=c^2dt^2-d\vec{x}^2$).

\section{Invariant equation of motion}
As we have shown, the invariant vector $\left(\frac{d\vec{p}}{dt}\right) _{\parallel v} + \gamma_{v} \left(\frac{d\vec{p}}{dt}\right) _{\perp v}$  is equal to $m\,d\vec{v}_R/d\tau$, measured in the rest frame $R$ of an accelerating body. However, in the frame $R$,  the classical relation between a force and acceleration is valid:
\begin{equation}
\vec{F}_R= m\frac{d\vec{v}_R}{d\tau},
\label{NEWT}
\end{equation}
 where the force $\vec{F}_R$ is a function depending on some features of the force source established in the frame $R$. 
For example, for a charge $q$ we have $\vec{F}_R=q\vec{E}_R$, where $\vec{E}_R$ is an electric field described by Maxwell's equations. Certainly, 
once the function $\vec{F}_R$ is determined in the frame $R$, it is an established \emph{proper} quantity for a given physical situation.  As such, $\vec{F}_R$ is simply invariant and can be then used in \emph{any} system of reference. Taking the force $\vec{F}_R$  to describe motion in an arbitrary frame $S$ means that we use one and the same {numerically invariant} function $\vec{F}_R$, in  which merely its \emph{arguments} (space-time coordinates) should be transformed  from $R$ to a given system of reference $S$: $\vec{F}_R (\tau ,\vec{x}_R)\equiv \vec{F}_R [\tau (t,\vec{x}),\vec{x}_R(t,\vec{x})]\equiv\vec{F}_R(t,\vec{x})$.

Inserting the relation (\ref{prp'}) into (\ref{NEWT}) we get:  
\begin{equation}
\vec{F}_R(t,\vec{x})=\left(\frac{d\vec{p}}{dt}\right) _{\parallel v} + \gamma_{v} \left(\frac{d\vec{p}}{dt}\right) _{\perp v}.
\label{EQQ}
\end{equation}
This equation is valid for \emph{any} system of reference, so we have arrived at the desired \emph{truly invariant} relativistic equation of motion based on the invariant three-vectors.

The meaning of the proposed equation of motion in the form of (\ref{EQQ}) is very simple and intuitive. It is just the classical Newton's law (\ref{NEWT}) from the rest frame $R$, but written by means of the quantities used in a laboratory system of reference $S$. The function $\vec{F}_R$, originally  established
 in the frame $R$, appears to have  universal and fundamental significance because using it, one can  formulate the equation of motion in \emph{any} system of reference. It is understandable that the relevant features of the force source are to be established relative to the accelerating body, i.e. primarily in its instantaneous rest frame $R$, because solely the state of the source of force \emph{with respect to} the accelerating body should be meaningful.

It is easy to check that at the limit $v/c\to 0$, our equation of motion (\ref{EQQ}) becomes the classical Newton's 
second law of motion. So, the requirement of correspondence is fulfilled.

 In the context of the above argumentation  showing 
the clear intuitive meaning of the proposed equation of motion, it seems  that the conventionally used relativistic equation 
in the form of $\vec{F}=d\vec{p}/dt$, which is a direct copy of the classical one with the momentum redefined to the relativistic
 form, is conceptually less attractive than the proposed Eq.~(\ref{EQQ}). The standard equation of motion is not  truly  Lorentz invariant, but  merely Lorentz-covariant. Instead of $d\vec{p}/dt$, it seems to be 
 more useful  to regard the actual Lorentz-invariant (absolute) quantity 
$\left(\frac{d\vec{p}}{dt}\right)_{\parallel v}+\gamma_v\left(\frac{d\vec{p}}{dt}\right)_{\perp v}$ as the effect of the action of force. 
 Because it is equal to the well-defined proper force $\vec{F}_R$ established in the rest frame of a moving body, as an advantage we get an equation of motion (\ref{EQQ}) based on the Lorentz-invariant quantities. Additionally, because in Eq.~(\ref{EQQ}) we use one and the same function $\vec{F}_R$ in any frame,  we  avoid the problem of the relativistic force transformation leading   to a complex equation like (\ref{FF'}) or its four-dimensional counterpart. 

\section{Alternative version of classical electrodynamics}
According to the standard theory of electromagnetism, the equation of motion (i.e. its physically meaningful three-dimensional version) in a laboratory system $S$ is:
\begin{equation}
\frac{d\vec{p}}{dt}=q\vec{E}+q\vec{v}\times\vec{B},
\label{LF}
\end{equation}
where $\vec{E}$ and $\vec{B}$ are the electric field and magnetic induction in the frame $S$. If we pass to another frame $S'$, we get, according to Eq.~(\ref{pp'}):
\begin{equation}
\frac{d\vec{p'}}{dt'}=  \vec{F}_{_{\parallel V}} + \gamma_{_ V}\vec{F}_{_{\perp V}} -  \gamma_{_ V}\vec{v'} \times \left(\vec{\frac{V}{c^2}} \times \vec{F} \right),
\label{p'F}
\end{equation}
where, on the basis of Eq.~(\ref{LF}), $\vec{F}= q\vec{E}+q\vec{v}\times\vec{B}$. Substituting this explicit expression for $\vec{F}$ into Eq.~(\ref{p'F}) and expressing $\vec{v}$ by means of $\vec{v'}$ and $\vec{V}$ we obtain, after a bit of tedious algebra: 
\begin{equation}
\frac{d\vec{p'}}{dt'}=q\left(\vec{E}_{_{\parallel V}} +\gamma _{_ V}\vec{E}_{_{\perp V}}+\gamma_{_ V}\vec{V} \times \vec{B}\right)+q\vec{v'}\times \left( \vec{B}_{_{\parallel V}} +\gamma_{_ V}\vec{B}_{_{\perp V}}-\frac{\gamma_{_ V}}{c^2} \vec{V} \times \vec{E}\right)
\label{LF'}
\end{equation}
Subsequently, the electric and magnetic fields are defined  in the frame $S'$ as follows:
  \begin{equation}
\vec{E}'=\vec{E}_{_{\parallel V}} +\gamma _{_ V}\vec{E}_{_{\perp V}}+\gamma_{_ V}\vec{V} \times \vec{B},
\label{E'}
\end{equation}
\begin{equation}
\vec{B}'= \vec{B}_{_{\parallel V}} +\gamma_{_ V}\vec{B}_{_{\perp V}}-\frac{\gamma_{_ V}}{c^2} \vec{V} \times \vec{E}.
\label{B'}
\end{equation}
In this way, Eq.~(\ref{LF'}) appears to have the same \emph{form} as Eq.~(\ref{LF}):
\begin{equation}
\frac{d\vec{p'}}{dt'}=q\vec{E}'+q\vec{v'}\times\vec{B}'.
\label{LF'2}
\end{equation}
However, although Eq.~(\ref{LF'2}) is similar to Eq.~(\ref{LF}), the electromagnetic fields in each frame are completely different, numerically. It is reminiscent of the fact that the commonly used relativistic law of motion is only Lorentz-covariant, but not truly invariant.

In our opinion, the use of the standard equation of motion (\ref{EQ1}) and the above procedure for transforming this equation are conceptually rather inconvenient
 (''mixed'' electromagnetic fields that are different in different systems of reference). Instead of this, we propose to use the truly invariant law of motion (\ref{EQQ}). In the case of electromagnetism, 
this equation in a laboratory frame $S$ states explicitly that:
\begin{equation}
\left(\frac{d\vec{p}}{dt}\right)_{\parallel v}+\gamma_v\left(\frac{d\vec{p}}{dt}\right)_{\perp v}
= q\vec{E}_R.
\label{EM1}
\end{equation}
The vector $\vec{E}_R$ is the well-known electric field measured in the instantaneous rest frame of accelerating charge $q$. The arguments of  $\vec{E}_R$ (which is originally expressed by means of the quantities $(\tau,\vec{x}_R)$ used in $R$) should be expressed through the Lorentz transformation by $(t,\vec{x})$, used in the frame $S$, so that we get $\vec{E}_R (\tau ,\vec{x}_R)\equiv \vec{E}_R [\tau (t,\vec{x}),\vec{x}_R(t,\vec{x})]\equiv\vec{E}_R(t,\vec{x})$. If we pass to the frame $S'$, the equation of motion in this frame is:
\begin{equation}
\left(\frac{d\vec{p'}}{dt'}\right)_{\parallel v'}+\gamma_{v'}\left(\frac{d\vec{p'}}{dt'}\right)_{\perp v'}= q\vec{E}_R
\label{EM1'}
\end{equation}
and it remains merely to express the former arguments of $\vec{E}_R$ $(t,\vec{x})$  by  the primed ones $(t',\vec{x'})$ (or directly, the original variables $(\tau,\vec{x}_R)$ by the primed variables $(t',\vec{x'})$). 

Compared to the standard approach, the simplicity obtained thanks to the truly invariant equation of motion is evident. No transformation of the electromagnetic vectors is required. In any frame, we use one and the same function, $\vec{E}_R$, in which we must only transform its arguments to a given system of reference. The mathematical and conceptual transparency of this approach is emphasized by the fact that the vector $\left(\frac{d\vec{p}}{dt}\right)_{\parallel v}+\gamma_v\left(\frac{d\vec{p}}{dt}\right)_{\perp v}$ is understood as the invariant (and in this sense absolute) measure of the action of force. What is more, Eq.~(\ref{EM1}) shows that instead of the full set of Maxwell's equations for fields $\vec{E}$ and $\vec{B}$ in laboratory frames, it is enough to formulate a law of ''electromagnetism''  only for the field $\vec{E}_R$ registered in the frame $R$.  The field $\vec{E}_R$ alone allows us to  write the equation of motion (\ref{EM1}) for \emph{any} laboratory system of reference.

On the basis of Eq.~(\ref{LF}), we can express $\vec{E}_R$ by means of the standard electromagnetic fields measured in the frame $S$, which leads to:
\begin{equation}
\left(\frac{d\vec{p}}{dt}\right)_{\parallel v}+\gamma_v\left(\frac{d\vec{p}}{dt}\right)_{\perp v}= q(\vec{E}_{_{\parallel v}}+\gamma_v\vec{E}_{_{\perp v}})+q\gamma_v\vec{v}\times\vec{B}.
\label{EM2}
\end{equation}
The right hand side of the above equation  explicitly shows us the dependence of $\vec{E}_R$ on the velocity $\vec{v}$ of accelerating charge $q$ (the fields $\vec{E}$ and $\vec{B}$ are independent of $\vec{v}$). Surprisingly, we have discovered along the way a so far unknown invariant three-vector in the realm of electromagnetism. Namely,
\begin{equation}
\vec{E}_{_{\parallel v}}+\gamma_v\vec{E}_{_{\perp v}}+\gamma_v\vec{v}\times\vec{B}
\end{equation}
is a Lorentz-invariant three-vector.
It is invariant simply because it is the proper field $\vec{E}_R$, but expressed by the quantities and variables used in an arbitrary frame $S$.

Let us note that if we introduce new quantities defined as follows:
\begin{equation}
\vec{\mathcal{E}}\equiv\vec{E}_{_{\parallel v}}+\gamma_v\vec{E}_{_{\perp v}},
\label{NE}
\end{equation}
\begin{equation}
\vec{\mathcal{B}}\equiv\gamma_v\vec{B},
\label{NB}
\end{equation}
we can rewrite Eq.~(\ref{EM2}) as:
\begin{equation}
\left(\frac{d\vec{p}}{dt}\right)_{\parallel v}+\gamma_v\left(\frac{d\vec{p}}{dt}\right)_{\perp v} 
=q\vec{\mathcal{E}}+q\vec{v}\times\vec{\mathcal{B}}.
\label{EM3}
\end{equation} 
It is another form of the invariant equation of motion (\ref{EM1}). As we can see, Eq.~(\ref{EM3}) together with the definitions (\ref{NE})-(\ref{NB}) differ from the standard Eq.~(\ref{LF}), containing the  ordinary  electromagnetic fields only by the factor $\gamma_v$. This shows that the truly invariant equation of motion can be achieved from  the standard covariant one merely at the expense of this slight modification.

It would be interesting to find some laws that govern the quantities $\vec{\mathcal{E}}$ and $\vec{\mathcal{B}}$ (instead of Maxwell's ones for the fields $\vec{E}$ and $\vec{B}$) and then use the Lorentz-invariant law of motion in the form of Eq.~(\ref{EM3}) to describe the behavior of accelerating charged particles.

\section{Discussion and conclusions}

It is well known that Galilean transformation performed in three-dimensional Euclidean space leaves the acceleration vector unchanged. 
In turn, the Lorentz transformation in four-dimensional pseudo-Euclidean space-time from all possible four-vectors does not change only the
 null four-vectors (examples are the world line of the light
 $(ct, \vec{x}(t))$, where $|\vec{x}(t)|=ct$ or the wave four-vector $(\omega/c, \vec{k})$ for the electromagnetic wave).  
The quantities that remain the same under a transformation of coordinates of reference systems have a status of absolute 
features of a given physical process. In Newtonian physics, acceleration is something absolute that pertains to a moving object, 
while in special relativity, the speed of light is absolute, which is reflected in the above null four-vectors. To this point, however, it was not 
realized that in the realm of special relativity, a universal quantity may be ascribed not only to the light propagation but also to any accelerating object. 
It is the three-vector 
quantity $\left(\frac{d\vec{p}}{dt}\right) _{\parallel v} + \gamma_{v} \left(\frac{d\vec{p}}{dt}\right) _{\perp v}$, which is shown in this paper to remain unchanged under the Lorentz transformation. It contains  no superfluous additions, but only relevant information about a body's motion. Thus, we avoid the problem of redundant tensors as artificial  candidates for representing the observer-independent physical reality and get instead the three-dimensional quantity that corresponds to the physical process in a natural way. The same refers to the invariant three-vector $\vec{F}_R$ representing a proper force and permitting us to write the truly Lorentz-invariant equation of motion (\ref{EQQ}) based on invariant three-vectors.

Concluding, we want to point out that the traditional way of elaborating physical phenomena only within the tensor formalism seems to be an incomplete approach. Unquestionably, the tensor approach allows us to grasp and elucidate many geometrical aspects of special relativity, and to express in a concise way otherwise complicated relations among physical quantities 'entangled' by the Lorentz transformation. However, neglecting 'old-fashioned'  three-vector treatment resulted in overlooking some important dynamical invariants which have been  brought to light in this paper.

Let us remark that our method to identify invariant vectors by starting from acceleration measured in the rest frame of moving 
object succeeds equally for any type of coordinatization, and so also for a coordinatization based on 'everyday' clock synchronization 
invented first by Tangherlini, \cite{TAN} later discussed by Mansouri and Sexl \cite{MS} and recently by Selleri who named the transformations following 
from it the "inertial transformations". \cite{SEL1}$^-$\cite{SEL3}  Among outstanding features of the inertial transformation, let us mention that they assume existence of an absolute speed (and rest) of a body and absolute simultaneity of events for different observers moving with a relative velocity.   It is out of the scope of this paper to discuss in more detail the respective formulas for invariant vectors that can be derived from the inertial transformations. The most important is that our work is in the same spirit as the attempts to formulate special relativity in a more intuitive manner where some elements of physical reality are treated as universal and therefore, on theoretical grounds, are maintained the same for any observer. 

\section{Acknowledgements}
I wish to thank P. Prawda for inspiring me to elaborate the issue presented in this paper.
I am grateful to anonymous reviewers for their helpful comments that improved the final version of the manuscript.

\end{document}